\documentclass{emulateapj}
\usepackage{apjfonts}

\shorttitle{Submillimeter Sources Resolved by SMA}
\shortauthors{Wang et al.}

\begin{document}

\title{SMA\footnotemark[1] Observations of GOODS~850-11 and GOODS~850-13 --- \\
First Examples of Multiple Submillimeter Sources Resolved by an Interferometer}

\author{Wei-Hao Wang\altaffilmark{2}, 
Lennox L.\ Cowie\altaffilmark{3},
Amy J.\ Barger\altaffilmark{4,5,3}, 
and Jonathan P. Williams\altaffilmark{3}}

\footnotetext[1]{The Submillimeter Array is a joint project between the Smithsonian Astrophysical 
Observatory and the Academia Sinica Institute of Astronomy and Astrophysics and is 
funded by the Smithsonian Institution and the Academia Sinica.}
\altaffiltext{2}{Academia Sinica Institute of Astronomy and Astrophysics, 
P.O. Box 23-141, Taipei 10617, Taiwan}
\altaffiltext{3}{Institute for Astronomy, University of Hawaii, 
2680 Woodlawn Drive, Honolulu, HI 96822}
\altaffiltext{4}{Department of Astronomy, University of Wisconsin-Madison, 
475 North Charter Street, Madison, WI 53706}
\altaffiltext{5}{Department of Physics and Astronomy, 
University of Hawaii, 2505 Correa Road, Honolulu, H  96822}

\begin{abstract}
We carried out extremely sensitive Submillimeter Array (SMA) 340 GHz continuum imaging on
two submillimeter galaxies (SMGs): GOODS~850-11 and GOODS~850-13.  The observations reach
sub-mJy rms sensitivities and, interestingly, resolve both sources into multiple, physically unrelated SMGs.
GOODS~850-11 is resolved into two sources at different redshifts. GOODS~850-13 is resolved into three sources,
two with different spectroscopic redshifts and one only with a photometric redshift. All the SMA sources have 
fluxes in the 3--5~mJy range and all are detected at 1.4~GHz. Three of them are 
detected by \emph{Chandra}, and one is a previously unknown X-ray SMG.  
This is the first time that single-dish SMGs are resolved into
multiple unrelated sources and also the first time that the SMA has discovered new SMGs.  
Our results show that identifications of SMGs at any wavelengths other
than the submillimeter itself can be misleading, since such identifications usually only
pick up one of the real counterparts. Using simulations that mimic our SCUBA and SMA 
observations, we find that the number of triple systems detected in our SMA survey
is much higher than that expected from the current best-determined number counts.
We tentatively attribute this to clustering. We also predict that ALMA will
find $\sim1/3$ of $>5$~mJy 850 $\mu$m SCUBA sources to be multiple systems.
Based on our SMA observations and simulations, we suggest that large samples of existing 
SMGs should be imaged with sensitive interferometric observations, even if the SMGs were 
previously thought to be securely identified.
\end{abstract}

\section{Introduction}

Since the first discoveries of distant submillimeter galaxies (SMGs; 
\citealp{smail97,barger98,hughes98,eales99}), tremendous progress has 
been made in understanding their nature and their role in galaxy evolution.  
With single-dish submillimeter surveys, we are now able to resolve 
approximately 30\% of the 850~$\mu$m background  into point sources 
with $S_{850~\mu \rm m}\gtrsim2$ mJy and constrain their number counts 
fairly well at this bright end \citep[e.g.,][]{coppin06}.  However, the low
resolution of single-dish telescopes and the associated large positional 
uncertainties make the identification and 
followup of these sources quite difficult.  Roughly 60\%--70\% of bright SMGs have 
counterparts in deep radio interferometric images 
\citep{barger00,ivison02,chapman03b} and their positions are known with 
subarcsec accuracy. Spectroscopic followup of the radio identified SMGs 
shows a redshift distribution between $z\sim1.5$--3.5 and that the SMGs 
dominate the total star formation in this redshift range 
\citep{chapman03a,chapman05}.  The recent advent of the Submillimeter Array 
\citep[SMA;][]{ho04} further helps to identify the radio-faint SMGs, and the 
redshift distribution has been extended to $z>4$ 
\citep{iono06,wang07,younger07,cowie09, younger09}.
In addition to the redshift identifications, 
detailed followup observations have been made to study the properties of the
SMGs in the X-ray, near-infrared, mid-infrared, and molecular line transitions 
(e.g., \citealp{alexander03a}, hereafter A03; \citealp{swinbank04,pope08,yun08,greve05,tacconi06}).

The followup studies of SMGs have been overwhelmingly focused 
on the brighter sources ($S_{850~\mu \rm m}\gtrsim5$~mJy), which can be 
easily detected by single-dish telescopes. The nature
of fainter SMGs that comprise the bulk of the submillimeter background has been much 
less explored.  Surveys of lensing cluster fields yield small samples of sub-mJy 
sources \citep{blain99,cowie02,knudsen08}. The number counts indicate 
that the background is dominated by sources with $S_{850~\mu \rm m}\sim1$~mJy 
and that the full resolution of the background requires detections of $\sim0.1$~mJy 
sources.  Stacking analyses statistically detect faint SMGs at 500 $\mu$m to 1.2~mm,
and various attempts have been made to study the redshift distribution of these faint SMGs
(e.g., Wang et al.\ 2006; \citealp{serjeant08,marsden09,penner10}).  However, the results
from these stacking analyses have not converged to a consistent picture.
A full understanding of the submillimeter background population 
will likely require next-generation interferometers, such as the 
Atacama Large Millimeter/Submillimeter Array (ALMA).

Fortunately, it is now possible to detect more typical submillimeter sources 
with the SMA.  The recent upgrade of the SMA to a 4~GHz bandwidth 
not only greatly boosts its continuum sensitivity, but it also makes the calibrations 
with fainter quasars easier than before.  This provides 
a first glance of what we may find in deep ALMA surveys and what issues 
may be present in current studies of SMGs.  In this letter we report new 
SMA 340~GHz continuum observations of two SMGs, GOODS~850-11 and  
GOODS~850-13 (\citealp{wang04}, hereafter W04), aka.\ GN12 and GN21
\citep{pope05} in the Great Observatories Origins Deep Survey-North 
(GOODS-N; \citealp{giavalisco04}).  Interestingly, both single-dish sources are 
resolved by the SMA into multiple physically unrelated galaxies. To our knowledge, 
these are the first examples of resolved, unrelated, multiple sources in the SMG population.

\section{Observations and Data Reduction}
The SMA observations of GOODS~850-11 and 13 were carried out on
2009 December 30 and 31, respectively.  Seven of the eight SMA antennas were
available in the compact configuration.  Two sidebands of 4 GHz each 
were centered at 334 and 346~GHz.  Titan was observed to provide flux 
calibration. The bright radio source 3C 273 was observed to provide passband calibration.
Quasars 0958+655 and 1642+689, which are 17.4 and 25.1~degree from GOODS-N,
respectively, were observed every 15~minutes for complex gain calibrations.  
The 225~GHz opacity was excellent, $\sim0.04$--0.05 the
first night and $\sim0.05$--0.07 the second night.  The averaged single-sideband 
system temperature was 380~K on both nights.  The effective integration was 
$\sim4.8$ and 7.2~hour per antenna on GOODS~850-11 and 13, respectively.
 
The calibration and data inspection were performed with the IDL-based Caltech 
package MIR modified for the SMA. Continuum data were generated by averaging
the spectral channels after the passband phase calibration. Both gain calibrators
were used to derive gain curves. However, we compared the results made with
just adopting one calibrator for consistency checks, and we did not find a systematic difference.
Flux calibrations were performed using data taken under conditions (time, hour angle, 
and elevation) similar to that of the flux calibrator.  The flux calibration error is
typically within $\sim10\%$ with this method.  

The calibrated interferometric visibility
data were exported to the package MIRIAD for subsequent imaging and analysis.
The visibility data were weighted inversely proportional to the system temperature and 
Fourier transformed to form images. The ``robust weighting'' of \citet{briggs95} was also applied, 
with a robust parameter of 1.0, to obtain a better balance between beam size and S/N.  
The images were CLEANed around detected sources to approximately $1.5\times$ the 
noise level to remove the effects of the sidelobes.  The noises measured from
the CLEANed images are 0.72 and 0.68~mJy for the images of GOODS~850-11 and 13, respectively.
The synthesized beams are 
$2\farcs8 \times 2\farcs0$ and $2\farcs3 \times 2\farcs0$ in the images of GOODS~850-11 and
13, respectively. The images were corrected for the SMA primary beam response.
All our fluxes and flux errors are primary beam corrected.  
Source positions and fluxes were measured by fitting the image with point-source models
using the MIRIAD {\tt IMFIT} routine.

\section{Results}
\begin{figure*}[ht!]
\epsscale{1.0}
\plotone{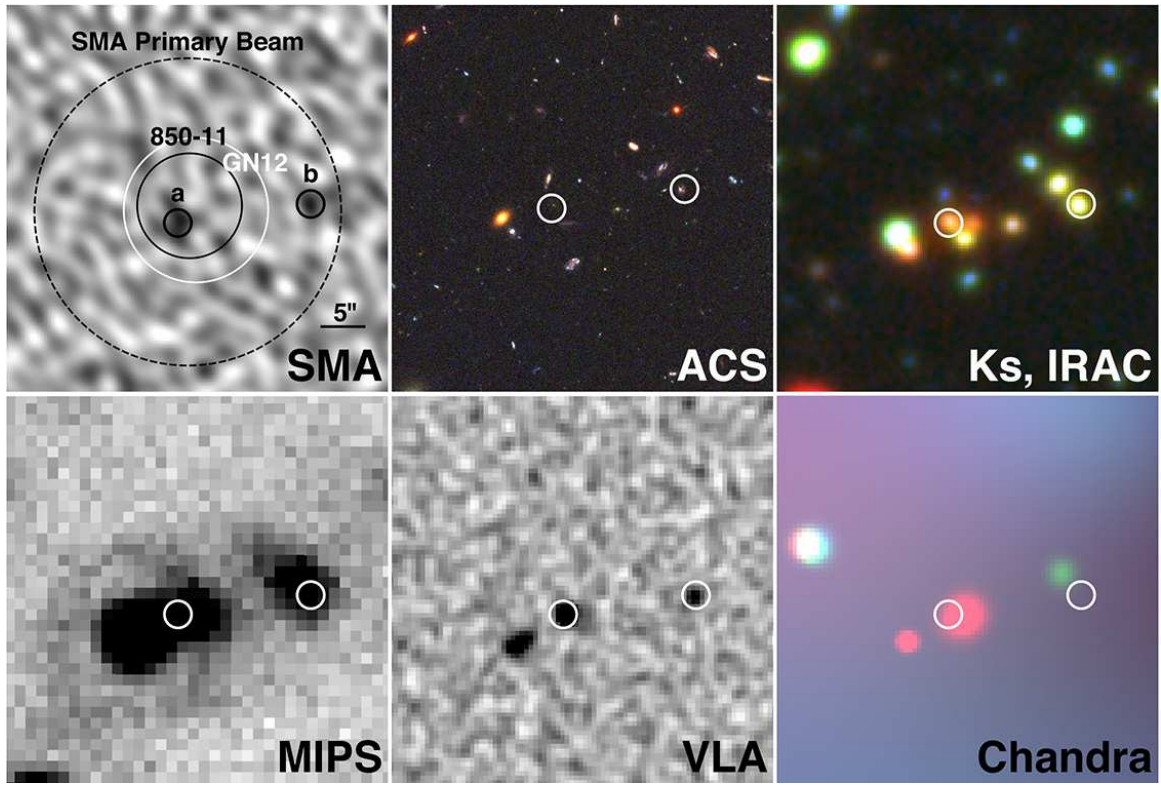}
\caption{Ultradeep multiwavelength images of GOODS~850-11. 
A false-color optical panel is made with \emph{HST} ACS F435W (blue), 
F606W (green), and F775W+F850LP (red) images. 
A false-color infrared panel is made with CFHT $K_S$ (blue; \citealp{wang10}), 
IRAC 3.6+4.5 $\mu$m (green), and IRAC 5.8+8.0 $\mu$m (red) images.
A false-color X-ray panel is made with adaptively smoothed \emph{Chandra}
4--8 keV (blue), 2--8 keV (green), and 0.5--2.0 keV (red) images \citep{alexander03b}.
The MIPS 24 $\mu$m image is from the \emph{Spitzer} GOODS Legacy Program 
and the VLA 1.4 GHz image is from \citet{morrison10}.
For visually uniform noise, the presented SMA image is uncorrected for the primary beam, 
which is shown with the very large dashed circles.  
Large solid circles show various SCUBA positions and the associated positional uncertainties determined by W04 (black) and 
Pope et al.\ (2005; white).  Small solid circles ($r=1\farcs5$) in all panels indicate the SMA sources centered at the SMA positions. 
All grayscale images have inverted scales.  North is up.
\label{fig_1}}  
\end{figure*}

\begin{figure*}[ht!]
\epsscale{1.0}
\plotone{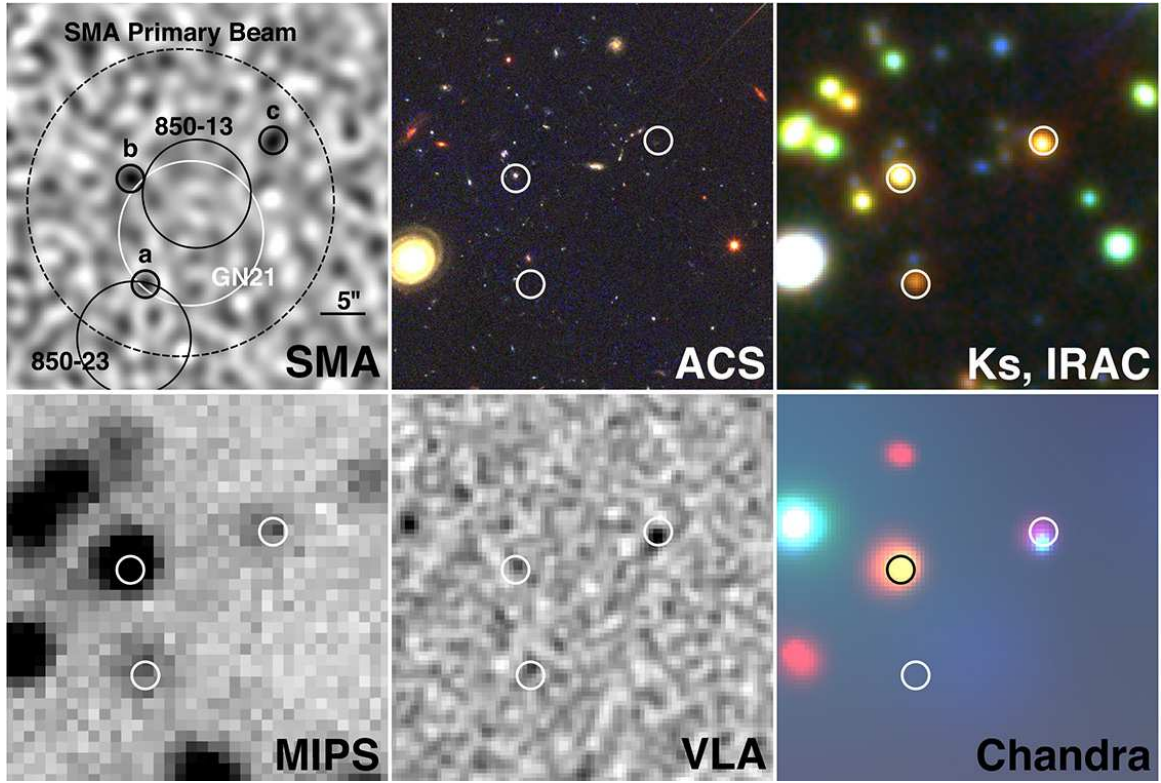}
\caption{Same as Figure~\ref{fig_1}, but for GOODS~850-13. \label{fig_2}}  
\end{figure*}

We present the SMA and multiwavelength images of GOODS~850-11 and 13 in 
Figures~\ref{fig_1} and \ref{fig_2}, their
SMA astrometry, long-waveband fluxes, infrared luminosities, and redshifts in 
Table~\ref{tab1}, and their optical to 
near-infrared spectral energy distributions (SEDs) in Figure~\ref{fig_sed}.  
The 1.4~GHz and 24~$\mu$m fluxes 
in Table~\ref{tab1} are adopted from \citet{morrison10} and the GOODS Spitzer Legacy Program DR2+ 
(M. Dickinson et al., in preparation), respectively, if the sources are detected. We carried out our own measurements 
for weakly detected sources. Below we describe the results.

\begin{deluxetable*}{lrrrrrrl}
\tablecaption{Basic Properties of the SMA Sources\label{tab1}}
\tablehead{\colhead{ID}  & \colhead{R.A.}  & \colhead{Dec.} & \colhead{$S_{\rm 340~GHz}$}
& \colhead{$S_{24~\mu\rm m}$} & \colhead{$S_{\rm 1.4~GHz}$} & 
\colhead{$L_{\rm IR}$\tablenotemark{a}} & \colhead{$z$\tablenotemark{b}} \\
& \colhead{(J2000.0)}  & \colhead{(J2000.0)} & 
\colhead{(mJy)} & \colhead{(mJy)} & \colhead{($\mu$Jy)} & \colhead{($L_\sun$)}
}
\startdata
GOODS~850-11a & 189.19137 & 62.24717 & $4.18\pm0.75$ & blended		& $109.6\pm5.3$ & $7.9\times10^{12}$	& 3.11 (2.76, 3.64)\\
GOODS~850-11b & 189.18324 & 62.24741 & $5.27\pm1.14$ & blended		& $32.6\pm7.3$   & $1.0\times10^{13}$	& 2.095 \\
GOODS~850-13a & 189.30845 & 62.19900 & $3.21\pm0.87$ & $93\pm7$ 	& $18.0\pm5.3$   & $6.1\times10^{12}$	& 3.46 (2.60, 4.50) \\	
GOODS~850-13b & 189.30944 & 62.20224 & $4.08\pm0.75$ & $216\pm7$ 	& $15.3\pm4.6$   & $7.8\times10^{12}$	& 3.157\\
GOODS~850-13c & 189.30002 & 62.20341 &$5.34\pm0.97$ & $52\pm7$ 		& $31.7\pm4.3$   & $1.0\times10^{13}$	& 2.914
\enddata
\tablenotetext{a}{Infrared luminosity derived from the SMA flux and the well-known negative $K$-correction in the submillimeter.
The conversion is $L_{\rm IR} = 1.9 \times 10^{12}~S_{850 \mu \rm m}~L_{\sun}/$mJy \citep[e.g.,][]{blain02}.}
\tablenotetext{b}{Redshifts with three significant digits are spectroscopic redshifts from Barger et al.\ (2008).  Redshifts with
two significant digits are photometric redshifts, with 99\% confidence ranges in the parentheses.}
\end{deluxetable*}

\subsection{GOODS~850-11}
GOODS~850-11 is known to be located in a region crowded with radio and X-ray sources (W04).
The SMA detected two sources, GOODS~850-11a and 11b (see Figure~\ref{fig_1}), 
at significance levels of 5.6 and 4.6~$\sigma$, respectively. 
In W04 we found a SCUBA point-source flux of $S_{850~\mu\rm m}=10.8$~mJy 
and a total flux of 12.7~mJy in a $30\arcsec$ region. The sum of the SMA fluxes
(Table~\ref{tab1}) of the two sources is consistent with our SCUBA measurements. 
\citet[hereafter P06]{pope06} found an 850~$\mu$m SCUBA flux of 
8.6 ~mJy, and they considered GOODS~850-11a as an unambiguous identification 
\citep[see also][]{greve08,chapin09}. 
A03 also considered this source to be the X-ray counterpart of the SCUBA source.  
Our SMA measurements show that this identification only accounts for $\sim50\%$ of the 
SCUBA flux in this region.  The brightest 24 $\mu$m source in this field (a blue ACS galaxy) 
is also detected in the radio and X-ray, but not in the submillimeter.
It is at $z=2.004$ and is consistent with an AGN with warm dust.

\begin{figure}
\epsscale{1.0}
\plotone{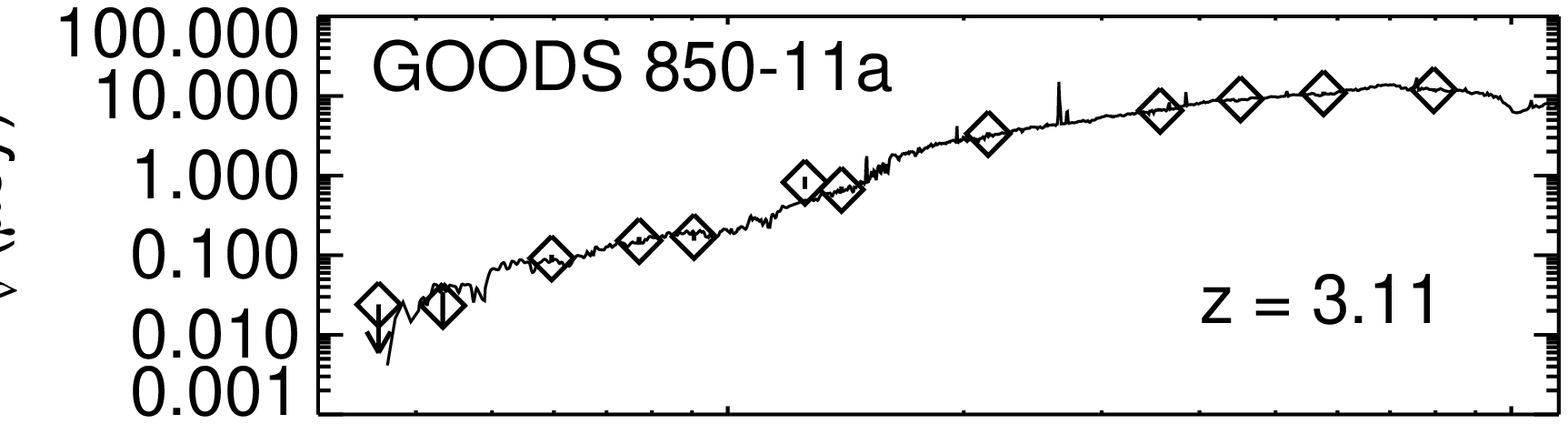}
\plotone{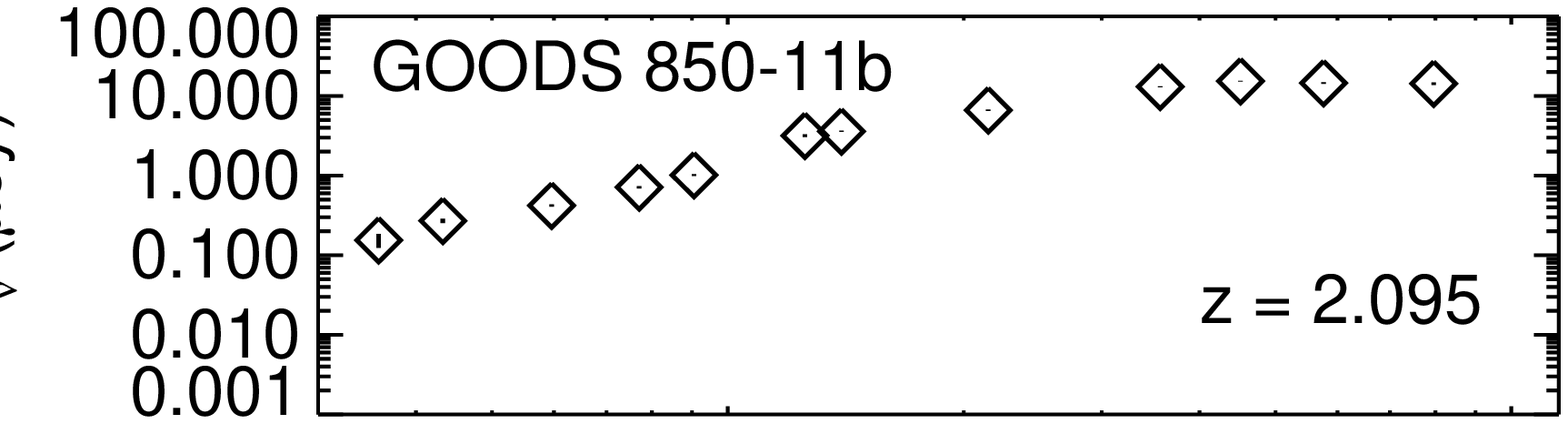}
\plotone{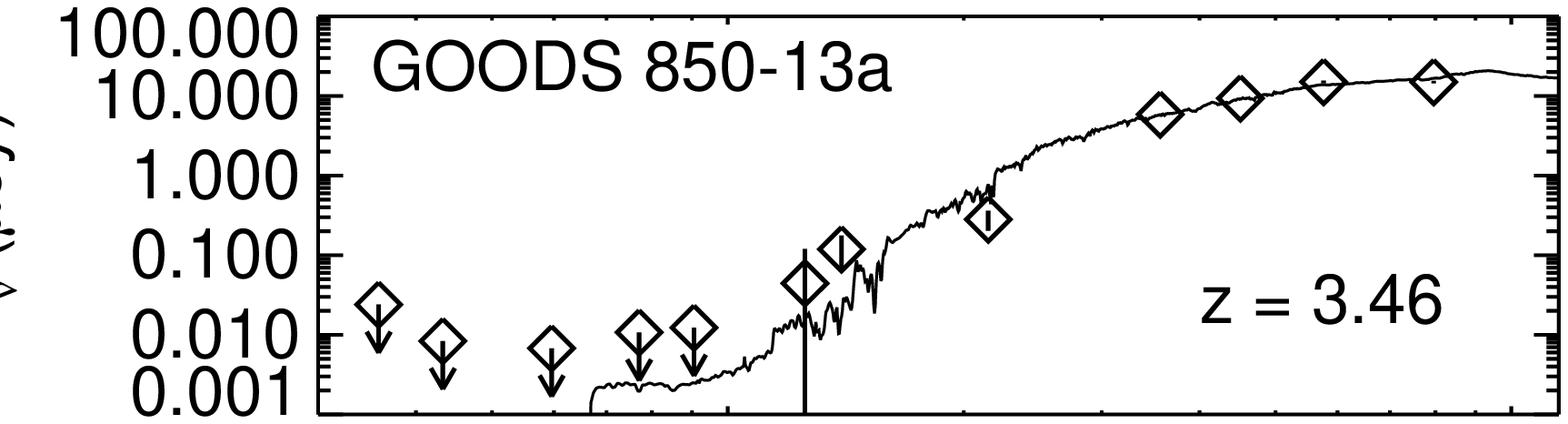}
\plotone{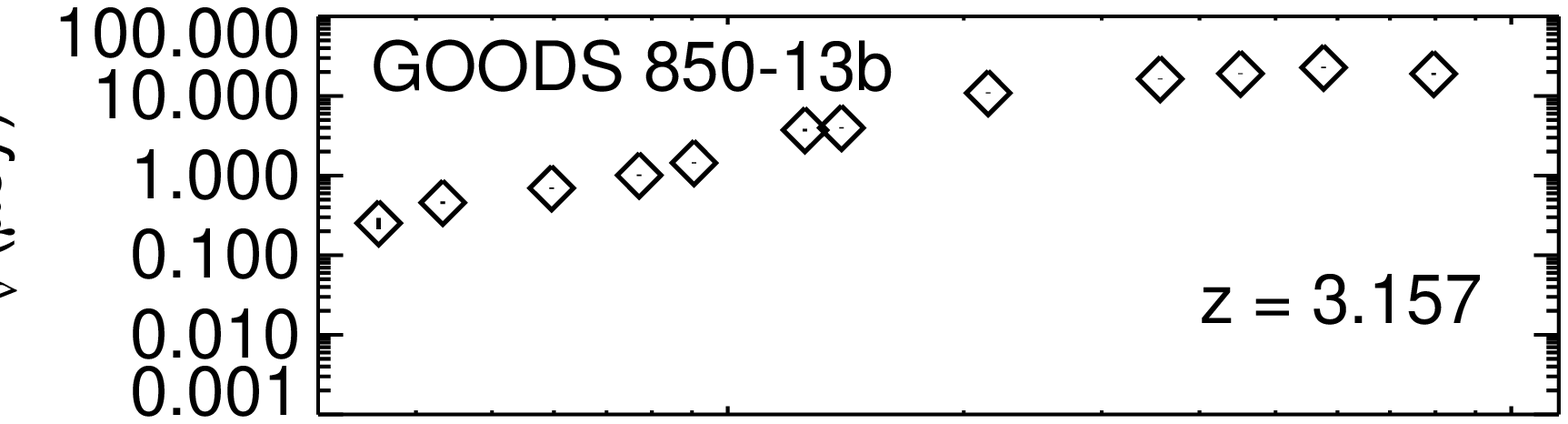}
\plotone{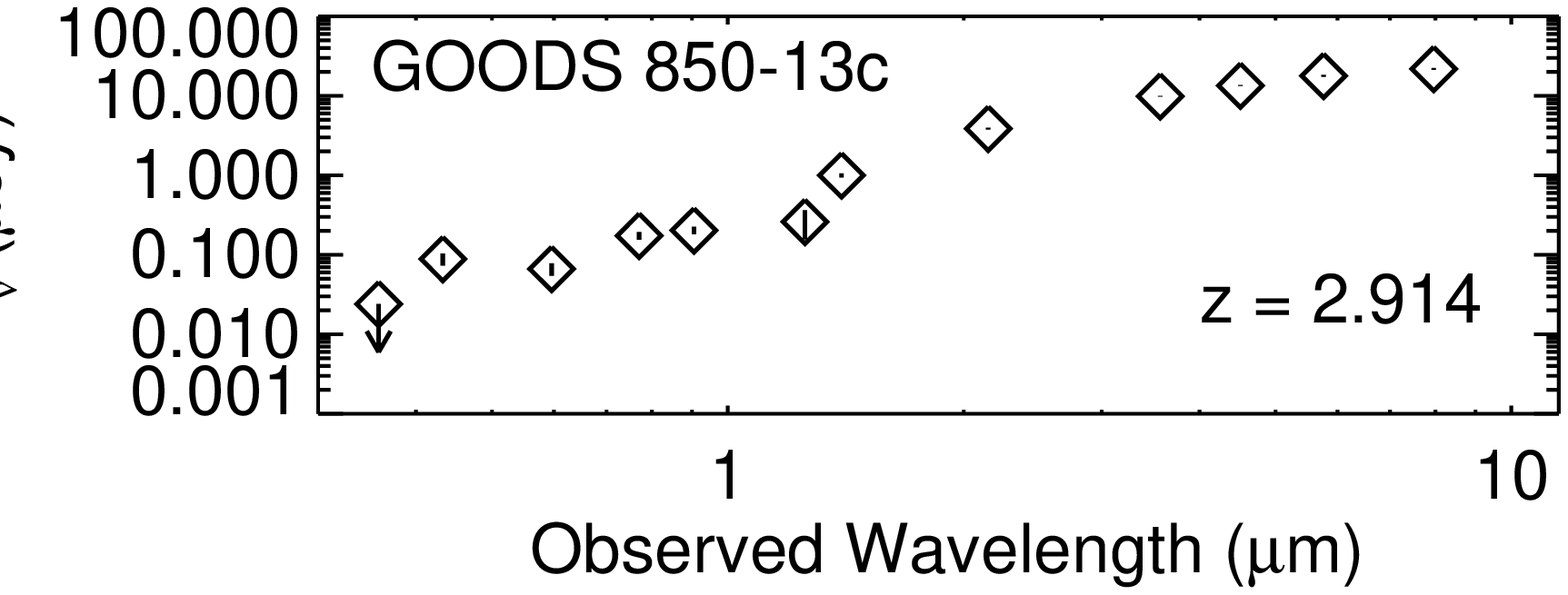}
\caption{Optical and infrared SEDs of the SMA sources. Symbols show observed SEDs. 
Best-fit SEDs for sources without spectroscopic redshifts are plotted with curves, and the corresponding redshifts
are the fitted photometric redshifts.
\label{fig_sed}}  
\end{figure}

\emph{GOODS~850-11a}---The source is detected at 1.4~GHz by the latest 
Very Large Array (VLA) imaging \citep{morrison10}.  It is also detected in the 
2~Ms \emph{Chandra} image (source number 239 in \citealp{alexander03b}).
Because it is severely blended with nearby sources in the MIPS 24~$\mu$m image, 
its 24~$\mu$m flux is highly uncertain.  
It does not have a spectroscopic redshift from our redshift survey \citep{barger08}.  
To estimate its redshift, 
we carried out photometric redshift fitting using the EAZY package \citep{brammer08}.  
We adopted the $U$-band flux 
from \citet{capak04}, the ACS fluxes from the GOODS v2.0 catalog, the $J$-band flux 
from a deep CFHT image that we will 
describe elsewhere, the \emph{HST} WFC3 F140W flux from 
A. Barger et al.\ (2010, in prep), and the $K_S$ to IRAC 
fluxes from Wang et al.\ (2010). We adopted the default set of SED templates of \citet{br07} plus a
dusty starburst model ($t=50$ Myr, $A_V=2.75$), all of which are provided in the 
EAZY package (see Brammer et al.\ (2008) for more details). 
These templates all include certain amounts of reddening, and we further reddened them 
by $A_V=0$, 0.5, and 1.0 with the extinction law of \citet{calzetti00} to account for very dusty sources.  
We allowed for any combinations of the above templates in the fitting.  
The photometric redshift result for GOODS~850-11a is $z=3.11$,
and the best-fit SED is presented in Figure~\ref{fig_sed}. 
This redshift agrees with the photometric redshift in P06, which is 3.1.

\emph{GOODS~850-11b}---The source is detected by the VLA at 1.4 GHz but not by the 
2~Ms \emph{Chandra} imaging.  
Similar to GOODS~850-11a, it is severely blended with nearby sources in the MIPS 24~$\mu$m 
image, and thus its 24~$\mu$m flux is highly uncertain. It has a spectroscopic redshift of 2.095 \citep{reddy06}.

\subsection{GOODS~850-13}
The SMA detected three components, GOODS~850-13a, 13b, and 13c, at significance levels of 
3.7, 5.5, and 5.5~$\sigma$, respectively. The significance level of GOODS~850-13a may not sound particularly high.
In a $36\arcsec$ SMA primary beam FWHM, there are, on average, $\sim18$ 3.6~$\mu$m sources 
and $\sim250$ synthesized beams.  For Gaussian noise, the probability of finding 
a $+3.7~\sigma$ noise peak
at the location of a 3.6~$\mu$m source is the associated Gaussian probability times 
18/250, which is $8\times10^{-6}$.  Thus, the coincidence with the 3.6~$\mu$m source
makes this detection much more secure.
We therefore conclude that GOODS~850-13a is a real detection.

In W04 we noticed an elongated morphology in our SCUBA image, and our
source extraction assigned the flux to two sources: GOODS~850-13 with 
$S_{850~\mu\rm m}=7.0$~mJy and
850-23 with $S_{850~\mu\rm m}=5.5$~mJy.  
The combined flux of the three SMA sources agrees excellently
with the combined flux of the two SCUBA sources.  
P06 adopted a different source extraction in their SCUBA image and
extracted only one source. They measured $S_{850~\mu\rm m}=5.7$~mJy, for which 
they identified GOODS~850-13a
as the counterpart.  Since then, 13a has been accepted as the 
counterpart in the literature \citep[e.g.,][]{greve08}. 
As was the case for GOODS~850-11, P06's identification
is only partly correct and can only account for $\sim1/4$ of the flux in this region.  
GOODS~850-13c was earlier considered to be the X-ray counterpart of the SCUBA 
source by A03 (see also \citealp{chapman05}).  This identification did not agree with
that in P06. Our result shows that both previous identifications are only partially correct. 
Even if we include both P06's (13a) and 
A03's (13c) identifications, we still do not have a full picture.

\emph{GOODS~850-13a}---This source is detected by MIPS at 24 $\mu$m
but not by the 2~Ms \emph{Chandra} imaging.  It shows a hint of weak radio 
emission in the VLA 1.4~GHz image of \citet{morrison10}. 
We measured its 1.4~GHz flux in the VLA image with Gaussian fitting in 
the AIPS {\tt IMFIT} routine.  Ideally one
would like to measure the radio flux at the best-fit SMA position.  However, 
given the relatively low SMA S/N and therefore the larger positional uncertainty, 
we decided to measure the radio flux at the local radio peak. 
The result (Table~\ref{tab1}) may be biased by positive noise spikes and 
therefore should only be considered as an upper limit.  

GOODS~850-13a is extremely optically faint (undetected by ACS, Figures~\ref{fig_2} and \ref{fig_sed}), and
it does not have a spectroscopic redshift. Its photometric redshift (based on photometry 
in the $K_S$ and IRAC bands) is 3.46.  This redshift is very similar to the spectroscopic 
redshifts of 13b and 13c, so we cannot rule out a physical association between 13a and one
of the other two.

\emph{GOODS~850-13b}---This source is bright at 24~$\mu$m but quite faint in the radio.  
We measured its 1.4~GHz flux (Table~\ref{tab1}) in the same way as for GOODS~850-13a,
so it should be considered as an upper limit. On the other hand, GOODS~850-13b is 
detected in the 2~Ms \emph{Chandra} image (source number 377 in \citealp{alexander03b}),
making it a new X-ray SMG that was previously unknown in the literature.  It has a 
spectroscopic redshift of 3.157 (Barger et al.\ 2008).

\emph{GOODS~850-13c}---This source is significantly detected by the VLA \citep{morrison10} 
and by \emph{Chandra}  (source number 369 in \citealp{alexander03b}), but it is relatively 
faint at 24~$\mu$m.  It has a spectroscopic redshift of 2.914 \citep{chapman05}.

\section{Discussion}
Since the commissioning of the SMA, it has been used for followup observations of SMGs, 
either for identification 
\citep[e.g.,][]{iono06,wang07,younger07,cowie09,younger09} or for morphology \citep[e.g.,][]{younger08}. 
This is the first time that the SMA has discovered new SMGs. The multiple 
SMA detections illustrate the limitations of identifying SMGs in any wavelength other than the submillimeter itself.  
Both sources had radio, 24 $\mu$m, and X-ray identifications in P06 and A03.  All of the previously proposed 
identifications are only partially correct; i.e., they are all legitimate SMGs, but the submillimeter fluxes and source 
numbers will be misinterpreted by up to a factor of 3.  If such cases are common, then 
our understanding of the SMG population is fundamentally flawed.  

The effects of multiple SMGs are commonly included in single-dish number counts simulations 
(e.g., \citealp{eales00,scott02,coppin06}; W04). However, its importance (relative to other effects such
as Eddington bias) has not been directly demonstrated by observations.
So far only sensitive millimeter/submillimeter interferometric observations can reveal the existence of 
multiple SMGs like GOODS~850-11 and 13, since existing single-dish telescopes are still severely 
confusion limited (at $>850~\mu$m) or noise limited (at 450~$\mu$m).  
The SMA surveys of \citet{younger07,younger09} imaged 15 SMGs selected at 1.1 mm.  They did not find 
evidence of multiple sources, consistent with the argument made by \citet{ivison07}
that multiple sources are rare. However, the SMA surveys of Younger et al.\ have 345 GHz rms
sensitivities of 1--2 mJy.  Even if there are secondary sources with $S_{\rm 345~GHz}\sim3$--5 mJy in their survey 
fields, such sources would not be easily detected.  
Our SMA survey is the first one that is deep enough to reveal such cases.  

After the SMA upgraded to the 4 GHz bandwidth, we observed five GOODS-N SMGs at similar depths 
(A.\ Barger et al., in preparation). Two of them are resolved into multiple sources and reported here. 
There is a third resolved source that will be reported by Barger et al.  It is unclear whether this source
is a merger or a physically unrelated pair.  Even if we exclude the third source, the frequency of multiple 
sources in our SMA sample still seems unusually high.
To better understand this, we performed Monte Carlo simulations that mimic our SCUBA and SMA surveys.  

We first created 1000 simulated SCUBA images using 
the differential counts in \citet{cowie02} and \citet{coppin06}, 
the ``true-noise'' map created in our GOODS-N SCUBA survey in W04, and the SCUBA beam in W04. 
We than extracted the simulated SCUBA sources. 
For each SCUBA source detected at $>5$ mJy and $>4~\sigma$ (the selection criteria for our SMA observations), 
we searched the input catalog for any $>3$ mJy (our SMA detection limit) sources within $17\arcsec$ 
(the SMA primary beam HWHM) of the SCUBA position.  
In the simulations the mean number of $>5$ mJy sources in a W04 SCUBA area is 
$11.4\pm3.9$, where the uncertainty is the dispersion in the 1000 realizations and is nearly Poissonian.  
This value is consistent with the W04 SCUBA observations (15 sources).  On the other hand, it is 
significantly larger than the cumulative count, as it is affected by blending and flux boosting.  
The measured counts take these effects out.  The mean numbers of double and triple systems 
are $1.29\pm1.33$ and $0.06\pm0.27$, respectively.

We had observed 10 of the 15 $>5$ mJy SCUBA sources 
with the SMA. The most recent five of the 10 SMA observations were deep enough to detect $>3$ mJy sources. 
The earlier SMA observations were shallower, so there was a selection bias against SCUBA sources with multiple 
counterpart candidates.  Given this bias and to be conservative, we only scale the above values of $1.29\pm1.33$ and $0.06\pm0.27$
by a factor of 10/15, rather than 5/15.  We thus expect to find $0.86\pm0.89$ double and $0.04\pm0.18$ triple
systems.  The probability of finding one triple system like GOODS 580-13 is only 4\%, inconsistent with the actual observations.  
Among the possible explanations, the most likely one is clustering of SMGs, which is not
included in the simulations.  This is plausible because the photometric redshift of GOODS~850-13a has a confidence
range (Table~\ref{tab1}) covering the redshift of 13b or 13c.  This can be tested this with future spectroscopic observations
in the near-infrared or in the millimeter.

In the same simulations, we increased the SMA detection limit to 4~mJy, and we found that the probability of multiple 
systems dramatically decreases to $\sim6\%$ .  This is consistent with the SMA 
survey results of \citet{younger07,younger09}.  By altering the details of the simulations, 
we also found that the above results are fairly insensitive to the following SCUBA and 
SMA observing strategies: (1) the source extraction (various detection thresholds), (2) flux 
measurement in the SCUBA map (simple aperture flux vs.\ optimally filtered flux using the beam), 
(3) the shape of the SCUBA sidelobes (determined by the secondary chopping), and 
(4) the decision on where to point the SMA (as long as it is within the SCUBA positional uncertainty).

We can use our simulations to predict the early results of ALMA identifications of
SCUBA sources.  We adopt the primary beam HWHM of $8\farcs5$ of ALMA 
at 340~GHz. In the ALMA early science phase we expect to have at least 10 antennae, 
and we can detect 0.5~mJy sources in roughly one hour.  
If we point ALMA at a $>5$~mJy SCUBA source, the 
probabilities of detecting double and triple SMGs within the primary beam will be 29\% and 6.5\%, respectively. 
The combined fraction of $\sim35\%$ is very high.  We therefore predict that multiple detections 
in early ALMA observations will be quite common.

At the beginning of this section, we raised the issue about incomplete identifications of SMGs
in the X-ray, 24~$\mu$m, and radio.  Based on our SMA observations and the above simulations, 
we believe that multiple systems and therefore incomplete identifications are common.
Thus, we suggest that large numbers of single-dish sources should be re-identified
with sensitive interferometric observations, even if the sources were previously thought 
to be securely identified.

\acknowledgments
We thank Y.-W.\ Tang, M.\ Gurwell, and the
SMA staff for the help in acquiring and reducing the data, 
and the referee for the review.  
We gratefully acknowledge support from the National Science
Council of Taiwan grants 98-2112-M-001-003-MY2 and 99-2112-M-001-012-MY3 (W.H.W.),
NSF grants AST 0709356 (L.L.C.) and AST 0708793 (A.J.B.),
the University of Wisconsin Research Committee with funds
granted by the Wisconsin Alumni Research Foundation (A.J.B.),
and the David and Lucile Packard Foundation (A.J.B.).

\end{document}